\documentclass[aps,prl,onecolumn,superscriptaddress,10pt,floatfix,amsmath,amssymb]{revtex4-2}

\usepackage{graphicx} 
\usepackage{dcolumn}
\usepackage{xcolor}
\usepackage{bm}
\usepackage{physics}
\usepackage{lipsum}  
\usepackage{hyperref}
\usepackage[normalem]{ulem}
\graphicspath{{images/}}

\begin{document}

\title{Evaluating the performance of a weak-field homodyne receiver in quadrature phase-shift keying optical communication}

\date{\today}

   \author{Silvia Cassina} 
   \affiliation{Como Lake Institute of Photonics, Dipartimento di Scienza e Alta Tecnologia, Università degli Studi dell'Insubria, Via Valleggio 11, I-22100 Como (Italy)}
   
   \author{Alex Pozzoli} 
   \affiliation{Como Lake Institute of Photonics, Dipartimento di Scienza e Alta Tecnologia, Università degli Studi dell'Insubria, Via Valleggio 11, I-22100 Como (Italy)}

     \author{Michele N. Notarnicola}
   \affiliation{Department of Optics, Palack${\rm \acute{y}}$ University, 17. Listopadu 12, 779 00 Olomouc, Czech Republic} 
   
   \author{Marco Lamperti}
   \affiliation{Como Lake Institute of Photonics, Dipartimento di Scienza e Alta Tecnologia, Università degli Studi dell'Insubria, Via Valleggio 11, I-22100 Como (Italy)} 
  
\author{Stefano Olivares}
   \affiliation{Department of Physics, University of Milan, and INFN Section of Milan, Via Celoria 16, 20133 Milan, Italy}   
   
   \author{Alessia Allevi}  \email{alessia.allevi@uninsubria.it}
   \affiliation{Como Lake Institute of Photonics, Dipartimento di Scienza e Alta Tecnologia, Università degli Studi dell'Insubria, Via Valleggio 11, I-22100 Como (Italy)}

\begin{abstract}
Quantum communication protocols require efficient detection schemes to maximize the information transfer rate between the sender and the receiver. To this aim, we have demonstrated that weak-field receivers, merging wave-like and particle-like features, can be considered as a valid alternative to already existing receivers, such as optical homodyne detection. To better emphasize the potential of our receiver, in this work we consider a proof of concept for quaternary communication based on coherent states with the same amplitude and different phase values. The encoding in phase requires a fine control of phase noise obtained through a feedback system. The results achieved in terms of mutual information and secret key generation rate encourage further increase of the alphabet towards an approximately continuous phase modulation.
\end{abstract}

\maketitle

\section{Introduction}
We are currently living what is called second quantum revolution, in which quantum resources offer an advantage over classical ones in various contexts. Among them, it is worth mentioning the field of quantum communication, which is aimed at
devising secure communication protocols between two parties, usually called Alice and Bob, in the possible presence of an eavesdropper, usually called Eve, who can be easily detected \cite{cariolaro}.
In the optical framework and in the context of continuous variables, the encoding of information to be shared involves the different degrees of freedom of light \cite{flamini,cozzolino}. Amplitude and phase of quantum states of light are particularly exploited thanks to the availability of receivers that are sensitive to these quantities \cite{arrazola,arrazola1,dimario}. In particular, interferometric schemes, such as optical homodyne detection schemes \cite{grosshans,diamanti}, in which photon-number-resolving (PNR) detectors are employed could be highly desirable since they give access to both wave- and particle-like properties of light \cite{cattaneo}. Quite recently, we have developed a PNR detector that is endowed with such properties and we have demonstrated that it could be exploited for continuous-variable quantum key distribution (CV-QKD) protocols \cite{oe24,IJQI25}. The proof-of-principle test has been performed by considering the benchmark case of a binary communication, that is encoding information in two coherent states with the same amplitude and a $\pi$-difference in phase \cite{izumi,muller,OE17}.
However, in a realistic scenario CV communication requires the use of a larger alphabet, involving a discrete number of elements belonging to two complementary variables.
Previous works have shown that a suitably shaped 16-state constellation embedded with probabilistic amplitude shaping 
can well approximate a continuous modulation \cite{IEEE24,Parente}.
This kind of encoding can be achieved 
by either quadrature amplitude modulation (QAM) or amplitude-phase-shift keying (APSK), that both require joint modulation of the amplitude and phase of the encoded coherent states \cite{leverrier,hirano,ghorai,kunz}. Crucially, when the encoding is performed in more complex alphabets, involving a discrete phase modulation \cite{notarnicola25}, it is highly required that the phase noise sources, both due to the laser source and the environment (air, acoustic vibrations, ...), are reduced as much as possible. 
As a fundamental test of the feasibility of joint amplitude-phase modulations, in this work we address the paradigmatic case of four-symbol encoding \cite{becerra,becerra1} and consider a compact experimental scheme, whose noise is monitored and partially reduced by means of a Red Pitaya device. We apply this strategy to a
quadrature phase-shift keying (QPSK) optical communication protocol \cite{Konrad}, where the encoded coherent states get same amplitude and a different, equally spaced, phase, showing that our receiver is capable of detecting an increase of the value of mutual information shared by the two communicating parties \cite{sych}. The experiment is performed by means of a receiver based on a pair of Silicon photomultipliers (SiPMs), which are PNR detectors endowed with photon-number resolution \cite{chesi19,cassina21}. The obtained results encourage the further enlargement of the encoding alphabet by introducing QAM
modulation to realize a 4$\times$4=16 constellation of coherent states.
\section{Theory} \label{sec:theory}
\subsection{M-ary coherent state constellations}
\begin{figure}
	\centering
	\includegraphics[width=0.4\textwidth]{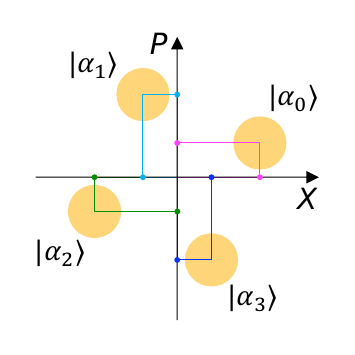}
	\caption{Sketch of the QPSK constellation analyzed in this work, with the reference phase $\phi_0=\pi/2M$ $(M=4)$.}
	\label{phase_space}
\end{figure}

As explained in the Introduction, in the case of discrete-modulation optical communications, 
classical information is encoded in a constellation of quantum CV states of radiation \cite{hager,thomas,biglieri,gaudenzi,sung}. Their arrangement in phase space is generally arbitrary. In order to improve the performance of the communication protocol, quantified in terms of mutual information, it makes sense to choose a constellation geometry that satisfies some symmetry \cite{cariolaro}, e.g. the Geometrically Uniform
Symmetry (GUS), being adopted in various quantum communication systems  \cite{NotarnicolaTesi}.
By definition, a constellation of $M$ states
\begin{equation}
	\{| \Psi_0 \rangle, | \Psi_1 \rangle, | \Psi_2 \rangle,..., | \Psi_{M-1} \rangle \}
\end{equation}
has a geometrically uniform symmetry when
\begin{itemize}
	\item{the $M$ states $| \Psi_k \rangle$ can be obtained by a single reference state $| \Psi_0 \rangle$ through the unitary transformation $| \Psi_k \rangle = \hat{U}^k | \Psi_0 \rangle$, being $k = \{0,1,...,M-1 \}$}\\
	\item{The symmetry operator is the Mth root of the identity operator, i.e. $\hat{U}^M = \mathbb{1}$}.
\end{itemize}
A practical example of GUS is provided by phase-shift-keying of order $M$ [PSK($M$)], where Alice prepares $M$ coherent states $| \alpha_k \rangle$ with the same amplitude $\alpha$ and different phases $\phi_k$, being equally spaced such that $\Delta\phi=\phi_{k+1}-\phi_k= 2\pi/M$ \cite{becerra,becerra1}. 
The uniform symmetry is then guaranteed by the phase-shift operator $U=\exp(2\pi i/M)$. 
Thus, the constellation can be expressed as $\{ |\alpha e^{i \phi_k}\rangle \}_{k=0,...,M-1}$, where
\begin{equation}
	\phi_k = \phi_0 + \frac{2 \pi k}{M} 
\end{equation}
and each state is sent with the same prior probability $q_k = 1/M$. In turn, the reference phase $\phi_0$ represents a free parameter whose value can be chosen to maximize the performance of the specific protocol under investigation. For instance, in the case of single-quadrature detection, e.g. homodyne detection of quadrature $\hat{x}$/$\hat{p}$, the most informative arrangement is obtained for the choice $\phi_0= \pi/2M$, such that the projections of the states on the $X$-axis and $P$-axis exhibit four distinct peaks as shown in Fig.~\ref{phase_space} for the QPSK case ($M=4$).
\subsection{Description of the optical communication protocol}
In this work, we will mainly focus on the QPSK constellation of Fig.~\ref{phase_space}, and consider application to an optical communication protocol over a pure-loss channel. That is, the sender (Alice) encodes a classical symbol $k=0,\ldots,M-1$ ($M=4$) on the coherent pulse $|\alpha_k \rangle$, being then sent to the receiver (Bob) through a quantum channel of transmissivity $T\le 1$. In turn, the received pulses still constitute a QPSK constellation $\{ | \sqrt{T} \alpha e^{i \phi_k}\rangle \}_{k=0,...,M-1}$ that preserves the GUS, albeit with rescaled signal amplitudes.
Unlike the conventional detection strategies based on homodyne measurements of either quadrature $\hat{x}$ or $\hat{p}$ \cite{Konrad, NotarnicolaTesi}, here we establish a suitable non-Gaussian receiver
based on the weak-field homodyne (WF) detection scheme presented in \cite{oe24}. According to this, the encoded coherent state $|\alpha_k \rangle$ is mixed at the balanced input beam splitter (BS) of a Mach-Zehnder interferometer with a mesoscopic local oscillator (LO) prepared in the coherent state $|z \rangle$, with amplitude $z>0$. At the two outputs of the interferometer Bob measures the conditional photon-number distributions
\begin{equation} \label{skellam}
	P_n(\mu_k|\alpha_k) = \exp(-\mu_k) \frac{\mu_k^n}{n!},
\end{equation}
by SiPM-based PNR detectors,
where $\mu_k$ assumes different expressions 
for either the reflected or the transmitted BS branch, i.e.
\begin{eqnarray}\label{meanvalues}
	\mu_k^{(t)} &=& \frac{1}{2} \left(T \alpha^2 + z^2 + 2 \xi \sqrt{T} \alpha_k z \right) \\
	\mu_k^{(r)} &=& \frac{1}{2} \left(T \alpha^2 + z^2 - 2 \xi \sqrt{T} \alpha_k z \right)
\end{eqnarray}
$\xi \in [0,1]$ being the spatial overlap between signal and LO, namely the interference visibility.\\
The present scheme allows WF detection of quadrature $\hat{x}$, that yields both information about the wave- and particle-like features of the probed field \cite{notarnicola25}, whereas WF measurement of $\hat{y}$ can be obtained by the same setup, provided that a $\pi/2$ phase-shift of the LO beam is performed. However, for the evaluation of the communication protocol in our case, it suffices to consider only WF $\hat{x}$ detection, given the symmetry of the considered constellation at Alice's side.
As we have already proved in Refs.~\cite{oe24,IJQI25} for binary PSK (BPSK) protocols, corresponding to $M=2$, if Bob is endowed with the WF $\hat{x}$ receiver described above, having access to both the single PNR detector outputs $(n, m)$ on the transmitted and reflected arms of the output BS, the mutual information (MI) shared by the legitimate parties is equal to:
\begin{equation}\label{mutual}
	I(A; B) = H(B) - H(B|A) = H[p_{\rm WF}(n, m)] - \sum_{k=0}^{M-1} q_k H[p_{\rm WF}(n, m|\alpha_k)],
\end{equation}
where $q_k=1/M$ is the prior probability,
$H[p(x)] = -\sum_x p(x) \log_2 p(x)$ is the classical Shannon entropy of $p(x)$, and the conditional probability of Bob retrieving the outcomes $(n, m)$ given the state $|\alpha_k \rangle$ sent by Alice, is equal to
\begin{equation}\label{eq:PWFCOND}
	p_{\rm WF}(n, m|\alpha_k) = P_n\left( \mu_k^{(t)}, \alpha_k \right) P_m \left(\mu_k^{(r)}, \alpha_k \right) = \frac{ e^{- \mu_k^{(t)}- \mu_k^{(r)}}}{n!m!} \left(\mu_k^{(t)}\right)^n \left(\mu_k^{(r)}\right)^m,
\end{equation}
while the overall Bob's probability distribution reads
\begin{equation}
	p_{\rm WF}(n, m) = \sum_{k=0}^{M-1} q_k p_{\rm WF}(n, m|\alpha_k) .
\end{equation}

Equation~(\ref{mutual}) can be compared with the MI obtained in the standard scenario of homodyne detection by Bob, where
\begin{equation}\label{eq:MIHD}
	I_{\rm HD}(A; B) = \sum_{k=0}^{M-1} q_k H[p_{\rm HD}(x|\alpha_k)] - \frac{1}{2} \log_2 \left(2 \pi e \sigma_0^2\right),
\end{equation}
where $e$ is the Euler's number,
\begin{equation}
	p_{\rm HD}(x|\alpha_k) = \frac{1}{\sqrt{2 \pi \sigma_0^2}} \exp \left[ \frac{(x-2 \sigma_0 \alpha_k \sqrt{T} \cos(\phi_k))^2}{2 \sigma_0^2} \right],
\end{equation}
is the conditional homodyne distribution, and $\sigma_0^2$ is the shot-noise variance, hereafter exprressed in shot noise units, namely $\sigma_0^2=1$ \cite{NotarnicolaTesi}.

\begin{figure}
	\centering
	\includegraphics[width=0.8\textwidth]{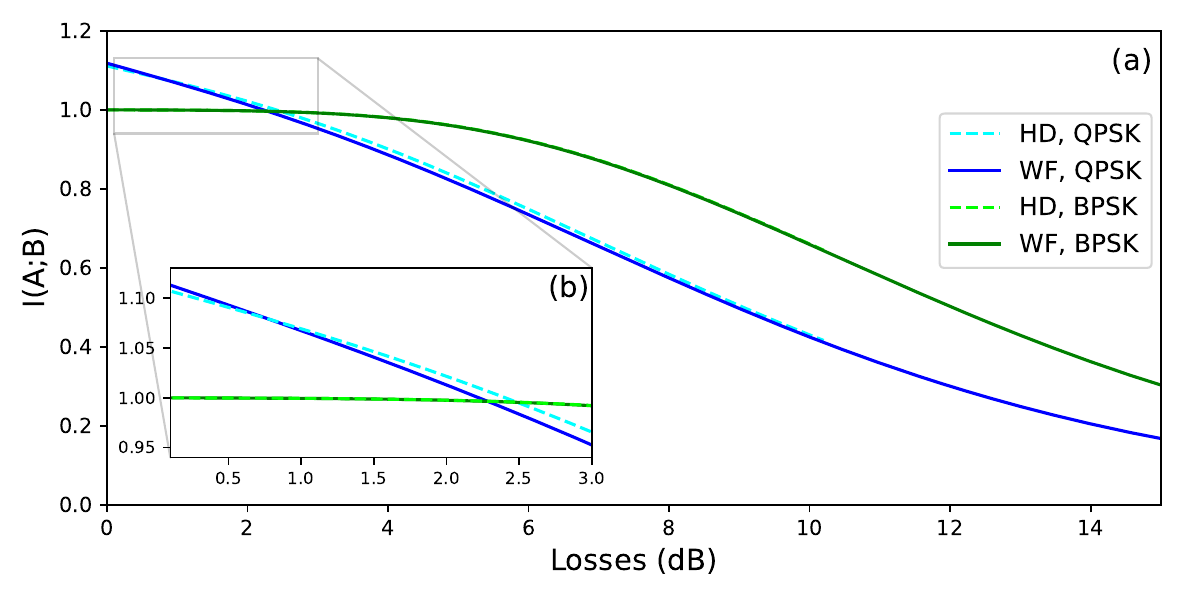}
	\caption{Mutual information as a function of the losses affecting the signal arm in the case of WF receiver (solid curves) and standard homodyne one (dashed curves). The blue curves refer to QPSK encoding (quaternary communication), while green ones to BPSK (binary communication). The inset displays an enlargement of the low-loss region. The values of the parameters are: $\alpha_{\rm MAX} = 2.04$, $z = 3.53$, and $\xi = 1$.}
	\label{MI_theor}
\end{figure}

Comparison between the WF and standard homodyne receivers for both BPSK ($M=2$) and QPSK ($M=4$) is observed in Fig.~\ref{MI_theor}, in which we plot the MIs~(\ref{mutual}) and~(\ref{eq:MIHD}) computed for a realistic choice of the signal and LO amplitudes 
as a function of the channel losses, equal to $-10\log_{10}(T)$ dB.
As it can be noticed, in both cases the results obtained with the WF receiver are almost indistinguishable from those achieved with the standard homodyne detection.
Moreover, the absolute values of the MI reached with the quaternary communication are larger than those achievable with a binary communication in the low-loss region, as shown in the inset of Fig.~\ref{MI_theor}. 
In fact, in the asymptotic limit $T \alpha^2 \gg 1$, the MI of a PSK($M$) protocol saturates to $\log_2 (M)$, corresponding to the maximum Shannon entropy of the source \cite{cover}, so that a realistic QPSK protocol can guarantee larger MI than BPSK for low channel losses and large enough signal energy. 
Hence, in the following we present the results of a proof-of-principle experiment by focusing on this range of losses affecting the signal.
\section{Proof-of-principle experiment}
\subsection{Experimental setup}
\begin{figure}
	\centering
	\includegraphics[width=0.8\textwidth]{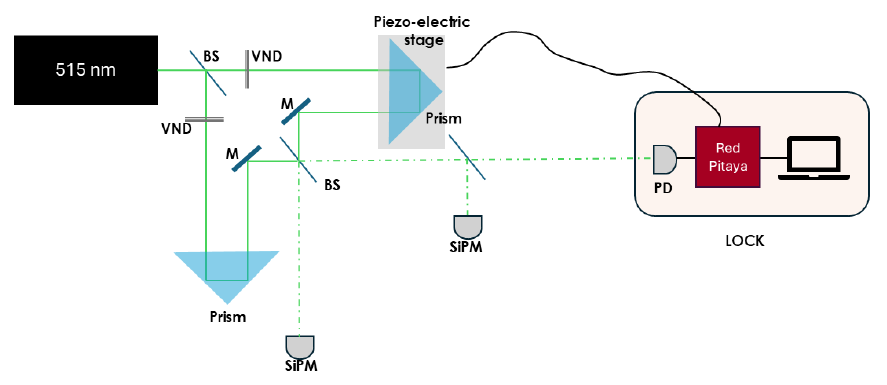}
	\caption{Sketch of the experimental setup. BS: beam splitter, VND: variable neutral density filter, M: mirror, SiPM: Silicon photomultiplier, PD: amplified photodiode.}
	\label{setup}
\end{figure}

The setup used to encode and decode information in 4 states while investigating the monitoring and control of the phase noise introduced by the light source and the environment is shown in Fig.~\ref{setup}. The picosecond pulses at 515 nm of a solid state laser operating at 0.5 MHz (mod. BDS-SM-515-FBC-101, Becker-$\&$-Hickl) are sent to a Mach-Zehnder interferometer, included in a box made of rigid polystyrene + sound-absorbing foam to reduce phase noise caused by air currents and acoustic noise. The reflected and the transmitted outputs of the input BS form the signal and the LO, respectively.
Two variable neutral density filters are placed in the two arms to finely tune the mean value of light separately. The right-angle prism in the signal arm is mounted on a micrometric stage to enable the optimization of the interferometer, while that in the LO arm is mounted on a piezoelectric movement connected to a feedback system to preserve the setup stability. In particular, we use a Red Pitaya device, connected to a computer, to monitor the fringe stability \cite{colavolpe} using a proportional integral (PI) control \cite{redpitaya,redpitaya1}. The input signal of the feedback system is given by the output of an amplified photodiode placed on a side output arm of the interferometer, as shown in Fig.~\ref{setup}.
At each output of the interferometer, the light is made divergent by a concave lens and selected by an iris to intercept only a very small portion of the interference pattern. Then, it is delivered to a SiPM detector through a multi-mode fiber (1 mm core fiber). The adopted model of SiPM (S13360-1350CS, Hamamatsu Photonics) is a matrix of 667 cells having a pixel pitch of 50 $\mu$m and operated in the Geiger-M$\rm \ddot{u}$ller regime with a common output. Each detector output is amplified by a fast-inverter amplifier (the amplification gain is equal to 17 dB), synchronously integrated by a boxcar-gated integrator over a 10 ns long gate centered around the peak, digitized and acquired at a repetition rate of 5 KHz.
The short integration window is adopted to reduce the contributions of dark counts and optical crosstalk, which generally affect the SiPMs. Indeed, by performing a dark measurement of $10^5$ samples, we experimentally get a value of about $0.3\%$ of dark counts for the two employed detectors, while the contribution of crosstalk effect is below $1\%$.
It is important to notice that the adopted boxcar-gated integrators represent the main limitation in terms of bandwidth (100 MHz) and noise, not allowing the reliable detection of light with mean number of photons lower than 0.5 \cite{OE25}. Reversely, the maximum detectable mean number of photons is equal to $\sim 15$. This means that, for smaller energy values or larger ones, a different acquisistion system should be used.
\subsection{Lock characterization}
\begin{figure}
	\centering
	\includegraphics[width=0.8\textwidth]{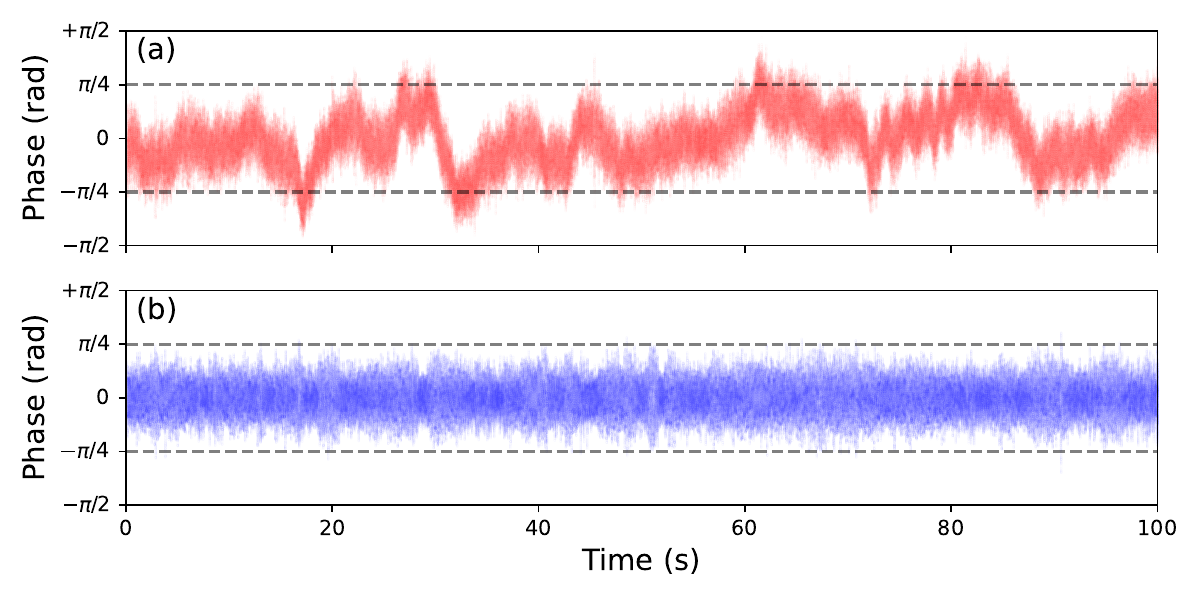}
	\caption{Temporal trace of the extracted phase in two conditions: (a) unlocked interferometer with box open; (b) fast lock engaged and box closed. Horizontal dashed lines represent a deviation of $\pm\pi/4$ with respect to the desired phase, corresponding to 0 rad in this figure.}
	\label{temporal_traces}
\end{figure}

\begin{figure}
	\centering
	\includegraphics[width=0.8\textwidth]{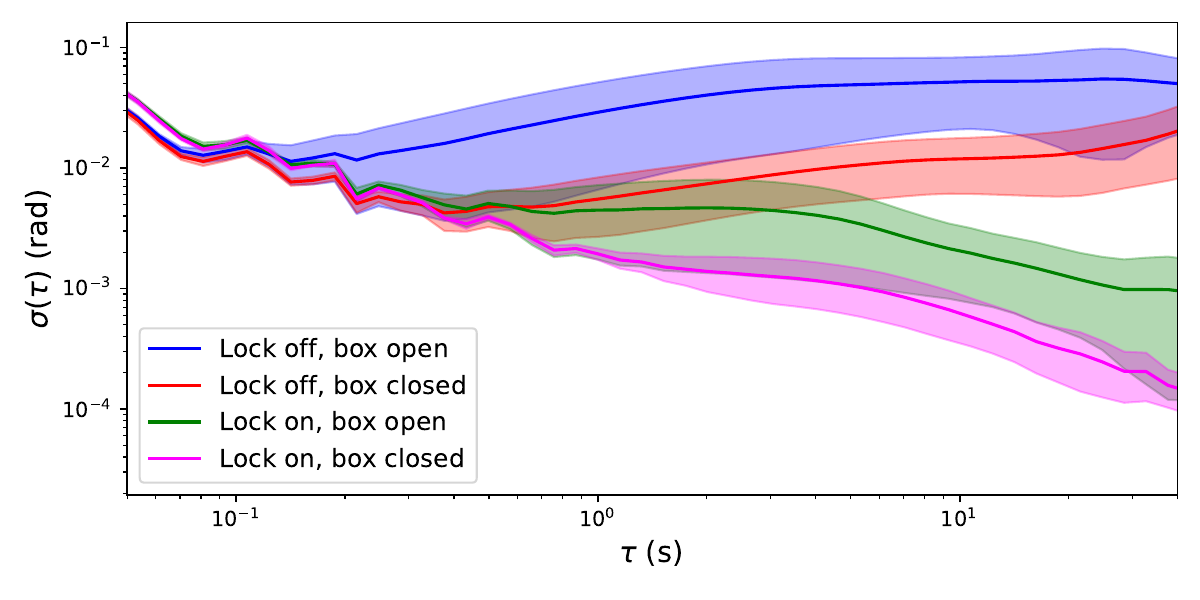}
	\caption{Graph of the overlapping Allan deviation curves for the four operating conditions: blue, lock off and box open; red, lock off and box closed; green, lock on and box open; purple, lock on and box closed. Solid lines are obtained as the averages of 10 independent measurements of Allan deviation, while shaded regions represent the corresponding standard deviation.}
	\label{Allan}
\end{figure}

\begin{figure}
	\centering
	\includegraphics[width=0.8\textwidth]{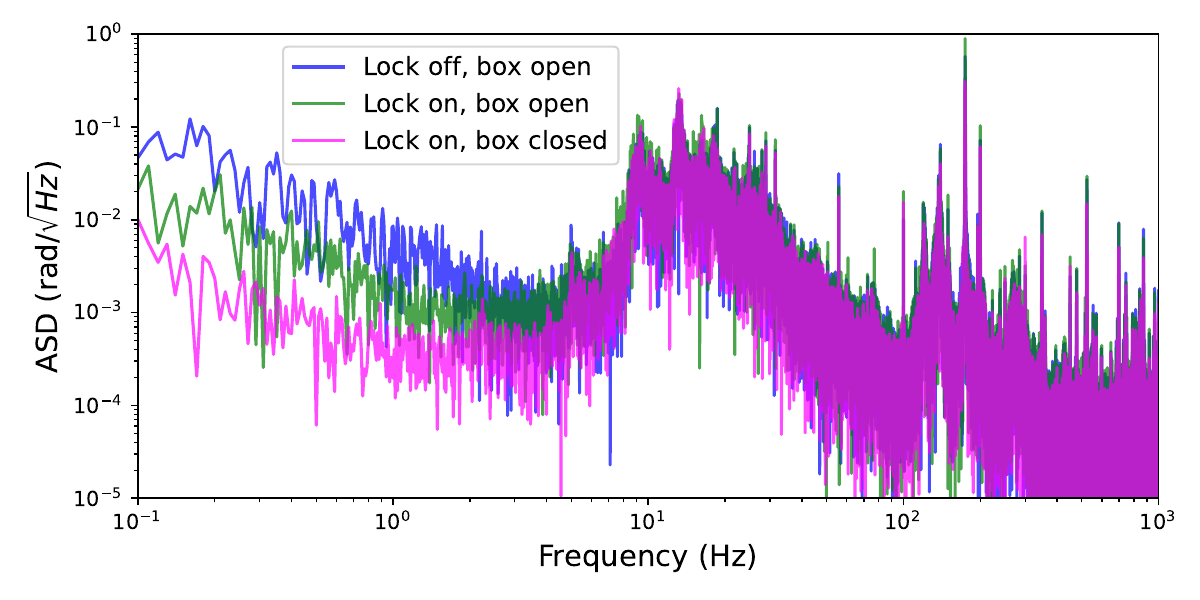}
	\caption{Graph of the ASD for three operating conditions: blue, lock off and box open; green, lock on and box open; purple, lock on and box closed.}
	\label{ASD}
\end{figure}

The primary purpose of the feedback loop is to eliminate slow drifts of the interferometric phase, which would otherwise prevent stable operation of the receiver over the timescales required for data acquisition. To characterize the lock performance, we extract the phase from the photodiode output signal by normalizing the fringe intensity between its minimum and maximum values and applying an arccosine transformation. The resulting phase time series is then analyzed both in the time domain, through the Allan deviation \cite{riley}, and in the frequency domain, through the amplitude spectral density (ASD).\\
The piezoelectric actuator employed in the feedback loop has a mechanical response limited to approximately 10 Hz, which sets an upper bound on the bandwidth of perturbations that can be actively corrected. We test two feedback configurations: a slow lock, in which only the integral gain is active with a value chosen to compensate for the drift, and a fast lock, in which a proportional term is added to minimize the residual phase noise within the actuator bandwidth. In addition to the active stabilization, we evaluate the effect of a closed box placed around the interferometer, designed to shield the setup from air currents and, to some extent, from environmental acoustic noise. Both feedback configurations are tested with the box either open or closed, resulting in a total of four operating conditions.\\
In Fig.~\ref{temporal_traces}
we show two representative temporal traces of the extracted phase, corresponding to the two extreme operating conditions. In panel (a) we present the unlocked interferometer with the box open: the phase exhibits both fast fluctuations and a pronounced slow drift. On the other hand, in panel (b) we show the phase recorded with the fast lock engaged and the box closed. In this case the drift is fully suppressed, while the residual fluctuations show a similar amplitude.\\
The effect of the different operating conditions on the phase stability is first assessed through the overlapping Allan deviation. The Allan deviation $\sigma(\tau)$ is a statistical measure, widely used in frequency metrology, that quantifies the stability of a signal as a function of the averaging time $\tau$ \cite{riley}. For a phase time series $\phi(t)$, it is defined as $\sigma^2(\tau) = \frac{1}{2\tau^2} \langle (\phi(t+2\tau) - 2\phi(t+\tau) + \phi(t))^2 \rangle$. When plotted on a log-log scale, the Allan deviation provides a direct visualization of the dominant instability mechanisms at each timescale: a decreasing $\sigma(\tau)$ indicates that the noise averages down with longer integration — as expected for uncorrelated perturbations — while an increasing $\sigma(\tau)$ signals the onset of drift or other long-term systematic effects. The minimum of the curve identifies the optimal averaging time, beyond which further integration tends to worsen rather than improve the phase estimate.\\
The Allan deviation measured with lock on or off and box open or closed in the four operating conditions is shown in Fig.~\ref{Allan}.
Comparing the curves reveals that neither the fast lock alone nor the box alone is sufficient to completely suppress the long-term fluctuations: only the combination of the fast lock with the closed box yields a monotonically decreasing Allan deviation over the full range of observed averaging times, indicating effective drift removal and reduced environmental coupling. The slow lock provides some degree of compensation of drifts, but its peformance is overall lower than the one of the fast lock, thus it is not considered anymore from now on.\\
The frequency content of the phase noise is further investigated through the ASD, which represents the square root of the power spectral density of the phase fluctuations and is expressed in units of rad/ $\sqrt{\rm Hz}$. The ASD provides a decomposition of the total phase noise into its spectral components, allowing one to identify the frequencies at which specific perturbation sources contribute most significantly. The ASD measured in the different operating conditions is shown in Fig.~\ref{ASD}.
Two prominent spectral features are observed: a broad peak around 20~Hz and a contribution near 200~Hz. The ASD corresponding to lock off and box closed is very similar to the one of lock off and box open, and is not plotted to improve the clarity of the figure. We attribute the former to mechanical vibrations transmitted through the floor to the optical table. This assignment is supported by the observation that deflating the pneumatic legs of the optical table leads to an increase of the 20~Hz peak, while the 200~Hz feature remains unchanged. The latter contribution is tentatively attributed to acoustic noise in the laboratory environment. While the box does not produce a significant reduction of this spectral feature — consistent with the fact that it is not airtight and is primarily designed to shield the setup from air currents rather than to provide acoustic isolation — its origin is consistent with the frequency range typical of environmental acoustic noise. When the fast lock is engaged, the ASD shows a clear reduction of the spectral density below 10 Hz, in agreement with the actuator bandwidth and consistent with the improvement observed in the Allan deviation at long averaging times.\\
Overall, the feedback system fulfills its primary function of eliminating the interferometric drift, while the residual phase noise is determined by environmental perturbations that lie beyond the actuator bandwidth. The RMS phase noise, evaluated from the temporal traces, decreases from 0.30 rad in the unlocked condition to 0.25 rad with the fast lock and closed box. The latter value is substantially smaller than the $\pi/2$ phase separation between adjacent states in the quaternary encoding, ensuring that the phase noise does not significantly affect the discrimination between the four coherent states. Hence, in the following we implement a proof-of-principle experiment in which the 4-state encoding is obtained with the system locked and taking into account the corresponding phase noise.
\subsection{Phase characterization}
\begin{figure}
	\centering
	\includegraphics[width=0.8\textwidth]{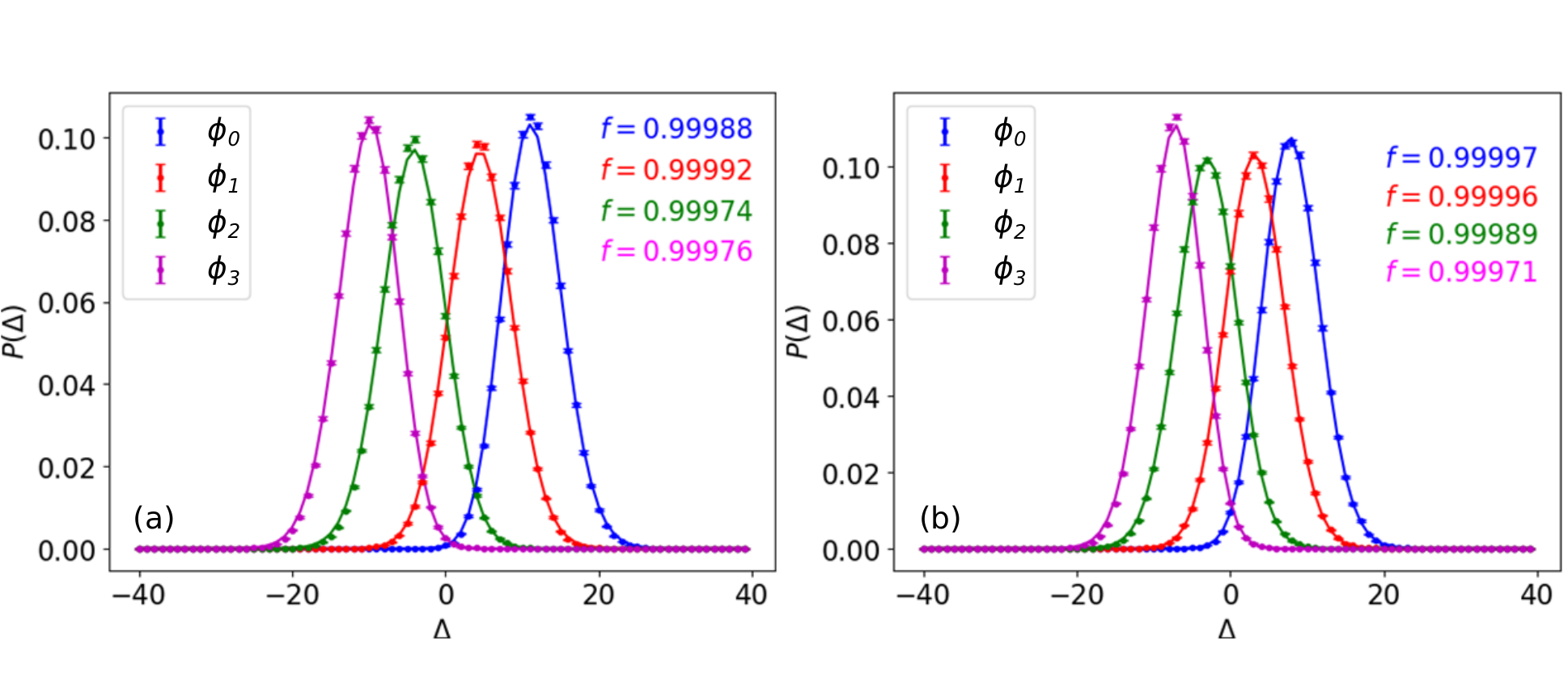}
	\caption{Photon-number difference for detected photons in the case of the QPSK encoding for two different choices of the mean value of signal: in panel (a) $\langle m_{\rm sig} \rangle = 4.13$, while in panel (b) $\langle m_{\rm sig} \rangle = 1.78$. Blue dots correspond to $\phi_0$, red dots to $\phi_1$, green dots to $\phi_2$, and purple dots to $\phi_3$. In both panels $\langle m_{\rm LO} \rangle = 12.5$. Colored dots: experimental data; colored solid lines: theoretical axpectations according to Eq.~(\ref{skellam}).}
	\label{fig:Skellam4phases}
\end{figure}

To prepare the QPSK encoding, we first align the detectors on the fringes corresponding to one of the states $| \pm \alpha \rangle$ and save a measurement in both conditions. In particular, for each experimental configuration, $2 \cdot 10^5$ consecutive pulses are acquired.
By applying the self-consistent method \cite{JMO} to the outputs of the detection chain, it is possible to reconstruct the statistical properties
of light. For instance, we can calculate the mean values of the maximum and minimum values of both states and estimate the imperfect overlap by using Eqs.(\ref{meanvalues}). Moreover, these values allow us to determine the expected intensity values corresponding to the desired phase shifts, corrected for the imperfect visibility. To set the phase shifts (and also to swap from $|+ \alpha \rangle$ to $|- \alpha \rangle$), we properly change the setpoint and sign of the PI controller. Since this procedure alone is not enough to obtain the exact desired values, small mechanical adjustments in the setup are also required for each state. As anticipated in Section~\ref{sec:theory}, in this work we aim at comparing the results obtained using $| \pm \alpha \rangle$, shifted by $\pi$, and the four states generated by applying to state $|+ \alpha \rangle$ the shifts $\phi_0 = \pi/8$, $\phi_1 = \pi/8 + \pi/2$,  $\phi_2 = \pi/8 + 3 \pi/2$, and  $\phi_3 = - \pi/8$.
The procedure described above is repeated for a fixed choice of the LO intensity, set roughly equal to $\langle m_{\rm LO} \rangle = 12.5$, and different values of the signal intensity. This variation is obtained by rotating the neutral density filter (VND in Fig.~\ref{setup}) on its arm, taking care to always work within the homodyne limit \cite{oe24,IJQI25}.
\begin{figure}
	\centering
	\includegraphics[width=0.8\textwidth]{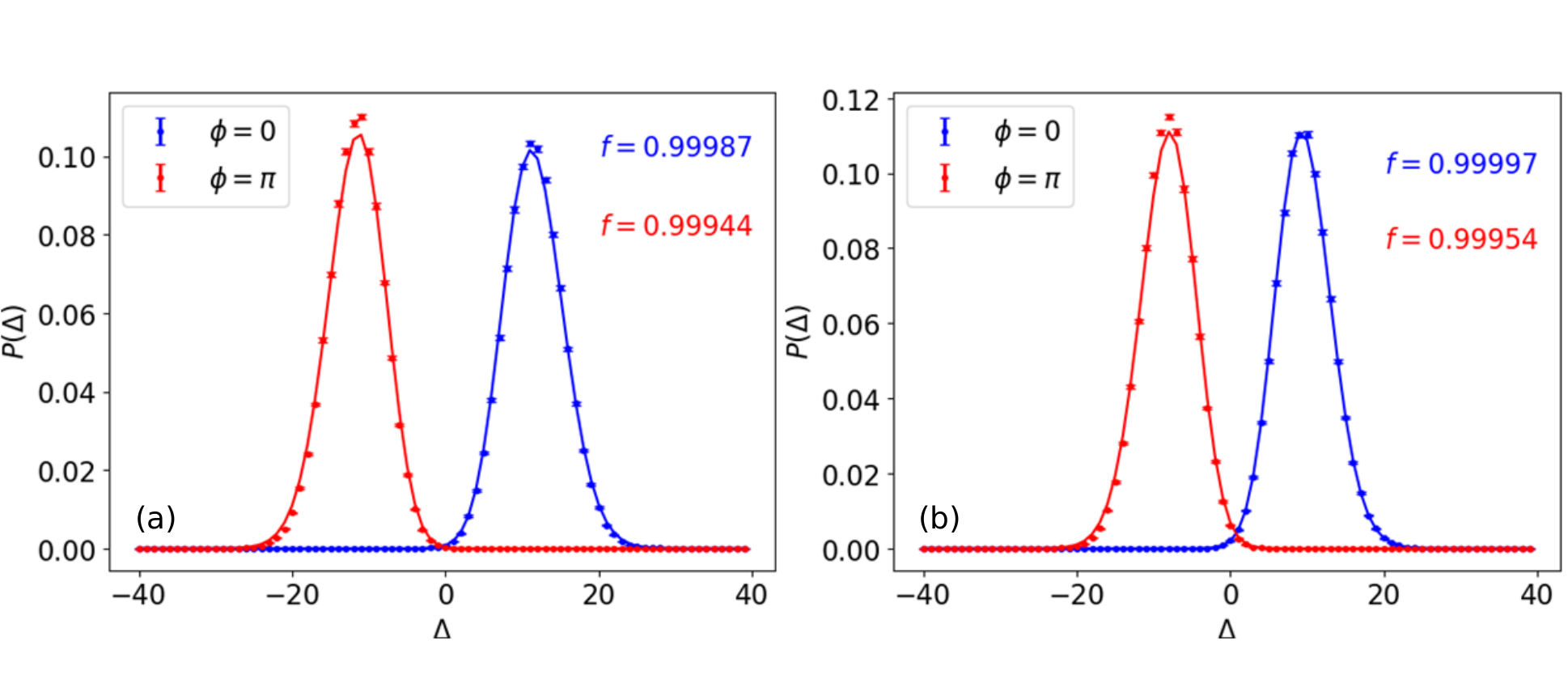}
	\caption{Photon-number difference for detected photons in the cases of $|+ \alpha \rangle$ (blue dots) and $|- \alpha \rangle$ (red dots) for two different choices of the mean value of signal: in panel (a) $\langle m_{\rm sig} \rangle = 4.13$, while in panel (b) $\langle m_{\rm sig} \rangle = 1.78$. In both panels $\langle m_{\rm LO} \rangle = 12.5$. Colored dots: experimental data; colored solid lines: theoretical axpectations according to Eq.~(\ref{skellam}).}
	\label{fig:Skellam2phases}
\end{figure}
\\
For instance, in the two panels of Fig.~\ref{fig:Skellam2phases} we show the reconstructed photon-number difference for detected photons in the case of $|+ \alpha \rangle$ and in the case of $|- \alpha \rangle$ for two different choices of the mean value of signal. We can notice that the lower the mean value, the larger the superposition and thus the harder the discrimination. Moreover, in both panels the distribution corresponding to state $|- \alpha \rangle$ (shown in red) is slightly higher than the distribution corresponding to state $|+ \alpha \rangle$ (shown in blue) as a consequence of the fact that the maximum intensities measured by the two SiPMs are not exactly equal because of slightly differences in the efficiency of the two detection chains. The theoretical expectations according to Skellam distribution in Eq.~(\ref{skellam}) are properly superimposed on each experimental distribution. The agreement between them is proved by the high values of the fidelity parameter, which is defined as $f= \sum_{m=0}^{\bar{m}} p_{\rm th}(m) p_{\rm exp}(m)$, in which $p_{\rm th}(m)$ and $ p_{\rm exp}(m)$ are the theoretical and experimental distributions, respectively, and the sum extends up to the maximum number of detected photons, $\bar{m}$, above which both $p_{\rm th}(m)$ and $ p_{\rm exp}(m)$ become negligible. Indeed, values of $f$ larger than 0.999 can be obtained.\\
The same analysis has been performed for the other 4 phases, as shown in the two panels of Fig.~\ref{fig:Skellam4phases} for two different choices of the mean value of signal.
In the figure, the experimental data are represented as dots, while the theoretical expectations are displayed as solid lines. Also in this case, according to the high values of the fidelity parameter, the experimental data are in agreement with theory. By comparing Figs.~\ref{fig:Skellam2phases} and \ref{fig:Skellam4phases}, it is evident that increasing the number of phase shifts the overlap between two curves increases. This behavior is more evident in panel (b), where the mean value of the signal is lower than in panel (a). While this translates into a difficulty in properly discriminating between two states \cite{PLA25}, on the other hand it increases the mutual information between Alice and Bob, i.e. the two parties communicating with each other.
\subsection{Evaluation of mutual information}
\begin{figure}
	\centering
	\includegraphics[width=0.8\textwidth]{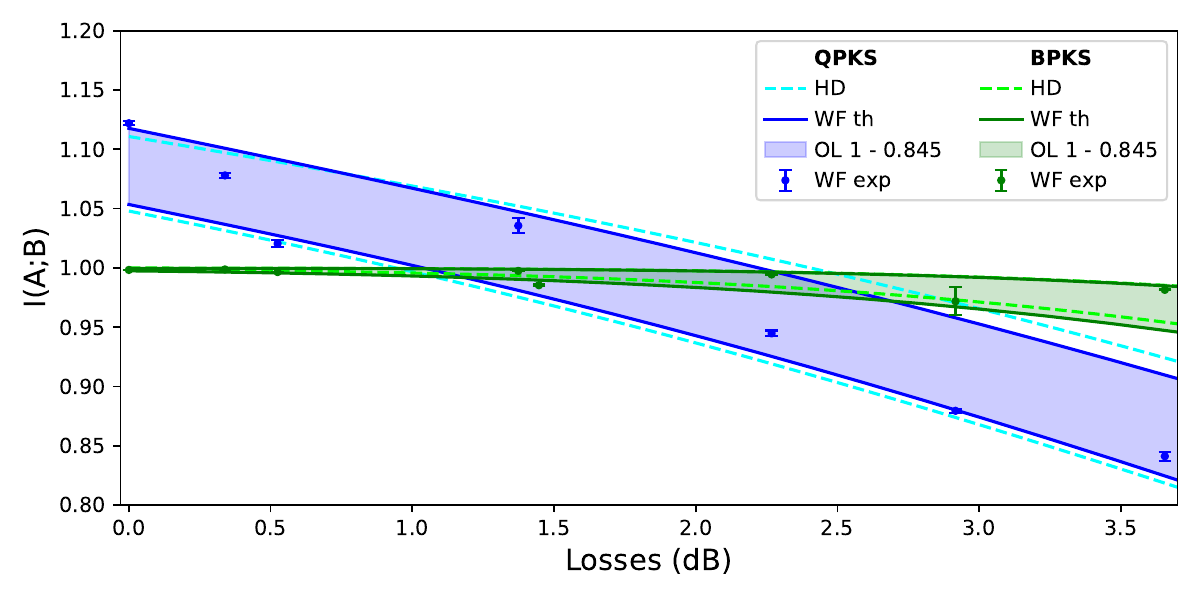}
	\caption{Mutual information as a function of the losses affecting the signal for BPSK (green) and QPSK (blue) encodings. Dots: experimental data, solid curves:  theoretical expectation in the case of WF receiver, dashed curves: theoretical expectation in the case of standard homodyne receiver. For both the theoretical expectations the upper curve refers to $\xi = 1$, while the lower curve to $\xi = 0.845$. The colored bands identify the region between these two conditions.}
	\label{MIexp}
\end{figure}

In order to prove this statement, we perform an analysis of the MI as a function of the losses affecting the signal, by keeping fixed the value of LO at the homodyne limit.
In particular, we directly compare the results of MI obtained by considering our WF receiver for BPSK (binary communication) and QPSK (quaternary one). At the same time, we present a comparison with the standard homodyne receiver.
The analyzed dataset is composed of seven different signal intensities, each of which has been acquired through four repetitions of $5 \cdot 10^4$ measurements. The average values of MI are shown in Fig.~\ref{MIexp} as a function of losses introduced in the signal arm.
The binary encoding is represented by the green dots, while 4-state encoding by the blue dots. We calculate the theoretical WF behavior, introducing the phase noise that characterizes our setup in the model, and assuming an imperfect overlap ranging from $\xi = 1$ (solid upper curve) and  $\xi = 0.845$ (solid lower curve).
We can observe that all the data fall within the bands delimited by these curves.
Moreover, for loss values lower than 2 dB, the MI obtained by the 4-state encoding is higher than the one obtained with the binary encoding, as already shown in Fig.~\ref{MI_theor}. Such a behavior is expected also in the case of standard homodyne detection, which is shown in Fig.~\ref{MIexp} as dashed curve for both the encodings. In addition, a direct comparison between the WF receiver and the standard homodyne one proves that, being all the other parameters the same, the photon-number resolution can be more advantageous in the case of imperfect overlap, at least in the explored region of low losses.

\section{Application to CVQKD}
The results obtained from the proof-of-principle experiment prove that classical information encoding over larger alphabets enables the possibility to reach higher values of MI even in the presence of single-quadrature detection schemes. This suggests to also address the application of the proposed QPSK-WF protocol for other quantum communication tasks, e.g. CVQKD, where, now, Alice's and Bob's goal is to share a common secure random key rather than transmitting a classical message \cite{NotarnicolaTesi}.
\begin{figure}
	\centering
	\includegraphics[width=0.8\textwidth]{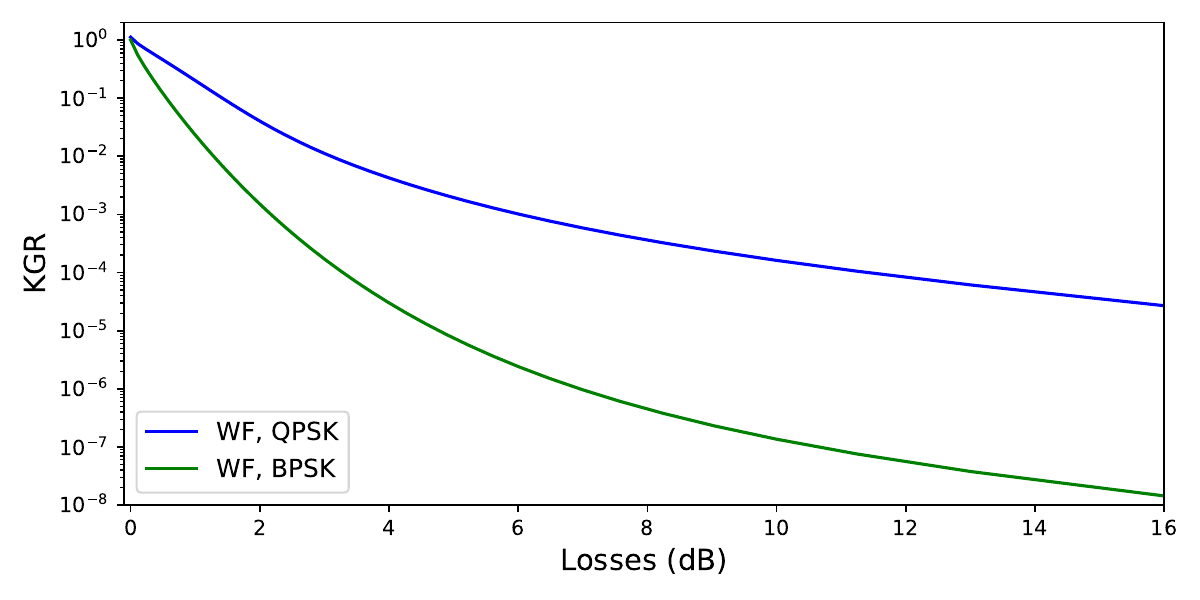}
	\caption{KGR as a function of the losses on the signal for BPSK (green curve) and QPSK (blue curve) modulation. The KGR is calculated in the case of collective attacks in the reverse reconciliation scenario.}
	\label{KGRteo}
\end{figure}
As a fundamental test, here we consider the same above-described optical communication setup, addressing CVQKD over a wiretap communication channel, where a third party, Eve, is performing beam-splitting attacks to collect the channel losses \cite{IEEE24}. In more detail, we assume that Eve has only access to the faction of the coherent signals $\{|\sqrt{1-T}\alpha_k\rangle\}_{k=0,\ldots,M-1}$ lost during channel transmission and that she performs collective attacks based on joint collective measurements over many stolen signals in the reverse reconciliation scenario \cite{oe24}. Under all these conditions, the KGR reads
\begin{equation}
	{\rm KGR} = I(A; B) - \chi(B; E),
\end{equation}
$\chi(B; E)$ being the Holevo information between Bob and Eve, namely the maximum information that can be extracted from the ensemble of eavesdropped signals according to quantum mechanics laws \cite{NotarnicolaTesi}, and equal to $\chi(B; E)= S(E) - S(E|B)$, where $S(E)$ and $S(E|B)$ are the unconditional (conditional) von Neumann entropies $S[\rho]=- {\rm Tr}[\rho \log_2 (\rho)]$ at Eve's side, respectively. After straightforward calculation, we find $S(E)=S[\rho_E]$, where $\rho_E= \sum_{k=0}^{M-1} q_k |\sqrt{1-T} \alpha_k\rangle \langle \sqrt{1-T} \alpha_k|$, and
\begin{equation}
S (E|B) = \sum_{n,m} p_{\rm WF}(n,m) \, S[\rho_{E|(n,m)}] \, ,
\end{equation}
with the $p_{\rm WF}(n,m)$ in Eq.~(\ref{eq:PWFCOND}), and  
\begin{equation}
\rho_{E|(n,m)} = \frac{1}{p_{\rm WF}(n,m)} \sum_{k=0}^{M-1} \, q_k \, p_{\rm WF}(n,m|\alpha_k) \, \left|\sqrt{1-T} \alpha_k \right\rangle \left\langle \sqrt{1-T} \alpha_k \right|
\end{equation}
being the corresponding conditional state received by Eve \cite{cattaneo,IJQI25}.

In Fig.~\ref{KGRteo} we compare the KGR in both the BPSK and QPSK cases as a function of the losses affecting the signal. It is well evident that, for the selected feasible values of signal and LO amplitudes, 4-state modulation outperforms the binary case at all values of losses, even in the regimes where the MI exhibits an opposite behavior. Incindentally, QPSK significanlty increases the KGR for high losses, thus reinforcing the need of larger coherent-state constellations in view of practical long-distance communication protocols.
Moreover, as already noticed in Ref.~\cite{IJQI25}, a realistic implementation of CVQKD protocols requires random-data encoding in datasets with large enough size so as to avoid unwanted KGR drop due to finite-size effects. This requirement is particularly important in high-loss regime, and futher implies the need of a fast acquisition system. 
Taking into account the recovery time of SiPMs (which is on the order of 5–10 ns) the maximum acquisition rate is set to 200 MHz, unless active quenching systems are exploited to further increase that rate \cite{lin}.\\
Thus, work is now in progress to develop a new detection chain capable of operating at a higher frequency and also less noisy than the one used in the proof-of-principle experiment.
Furthermore, to improve the effectivness of the receiver in the case of a larger alphabet it would be advisable to implement a double weak-field receiver \cite{peri} based on the use of four SiPMs instead of only two in order to perform for the first time WF double-homodyne detection, thus jointly probing the two orthogonal field quadratures $\hat{x},\hat{p}$.
This choice would make the receiver more versatile and fast, paving the way for efficient non-Gaussian quantum communication protocols in all energy regimes \cite{notarnicola25}.
\section{Conclusions}
In this work, we have performed a proof of feasibility of an optical communication protocol employing QPSK encoding of coherent states and a WF receiver. We reconstructed the shared MI between the users from the experimentally measured PNR statistics and demonstrated a larger amount of shared information with respect to the BPSK scheme for low channel losses even in the presence of single-quadrature detection. Furthermore, 
the obtained results are in agreement with those of a homodyne detection receiver, proving PNR detectors as a feasible alternative. The increase of the encoding alphabet, even though it requires precise control of the phase value, represents a crucial step towards the implementation of both optical communication and CVQKD, where QPSK is shown to significantly outperform BPSK.
Work is now in progress to improve the repetition rate at which the receiver can operate and to reduce the noise of the source as well as of the acquisistion system in order to make the WF detection system more reliable and effective.

\section*{Data Availability Statement}
The data that support the findings of this study are available from the corresponding author upon reasonable request.

\section*{Acknowledgments}
M.L. acknowledges the project ``Double weak-field homodyne receiver for the decoding of quadrature-amplitude-modulated coherent states'' supported by University of Insubria.
S.C. and A.A. acknowledge the support by PNRR D.D.M.M. 351/2022 and D.D.M.M. 737/2021.

\section*{Conflict of interest}
The authors have no conflicts to disclose.

\section*{Author contributions}
\textbf{Silvia Cassina:} Data curation (equal); Formal analysis (lead); Methodology (equal); Writing—Original draft (equal); Review and editing (equal). \textbf{Alex Pozzoli:} Data curation (equal); Formal analysis (equal); Writing—Original draft (equal); Review and editing (equal). \textbf{Michele N. Notarnicola:} Software (lead); Data curation (equal); Writing—Original draft (equal); Review and editing (equal). \textbf{Marco Lamperti:} Conceptualization (equal); Methodology (equal); Writing—Original draft (equal); Review and editing (equal). \textbf{Stefano Olivares:} Conceptualization (equal); Methodology (equal); Writing—Original draft (equal); Review and editing (equal). \textbf{Alessia Allevi:} Conceptualization (equal); Supervision (lead); Methodology (equal); Writing—Original draft (equal); Review and editing (equal).

\end{document}